\documentclass[fleqn,usenatbib,linenumbers]{mnras}
\usepackage{lineno}
\usepackage{comment}
\usepackage{enumitem}
\usepackage{amsfonts,amsmath,amssymb}

\usepackage{xspace}		 

\usepackage{newtxtext,newtxmath}
\usepackage[T1]{fontenc}

\DeclareRobustCommand{\VAN}[3]{#2}
\let\VANthebibliography\thebibliography
\def\thebibliography{\DeclareRobustCommand{\VAN}[3]{##3}\VANthebibliography}

\hypersetup{linkcolor=red,citecolor=cyan,filecolor=blue,urlcolor=magenta}
\usepackage{subfigure}

\def\P{{\rm P}}
\def\lappeq{\mathrel{\rlap{\raise.5ex\hbox{$<$}}{\lower.5ex\hbox{$\sim$}}}}

\usepackage{graphicx}	% Including figure files
\usepackage{amsmath}	% Advanced maths commands

\title[Lu et al.]{Reconstruction of luminosity function from flux-limited samples}

\author[Lu et al.]{
	Rui-Jing Lu,$^{1}$\thanks{E-mail: luruijing@gxu.edu.cn}
	Wen-Hao Chen,$^{1}$
	Wen-Qiang Liang$^{1}$
	and Cheng-Feng Peng$^{1}$
	\\
	$^{1}$Guangxi Key Laboratory for Relativistic Astrophysics, School of Physical Science and Technology, Guangxi University, Nanning 530004, People’s Republic of China\\
}

\begin{document}
\label{firstpage}
\pagerange{\pageref{firstpage}--\pageref{lastpage}}
\maketitle

\begin{abstract}
The properties of the progenitors of gamma-ray bursts (GRBs) and of their environment are encoded in their luminosity function and cosmic formation rate. They are usually recovered from a flux-limited sample based on Lynden-Bell's $c^{-}$ method. However, this method is based on the assumption that the luminosity is independent of the redshift. Observationally, if correlated, people use nonparametric $\tau$ statistical method to remove this correlation through the transformation, $L^{\prime}=L/g(z)$, where $z$ is the burst redshift, and $g(z)=(1+z)^{k}$ parameterizes  the underlying luminosity evolution. However, the application of this method to different observations could result in very different luminosity functions. By the means of Monte Carlo simulation, in this paper, we demonstrate that the origin of an observed correlation, measured by the $\tau$ statistical method, is a complex combination of multiple factors when the underlying data are correlated. Thus, in this case, it is difficult to unbiasedly reconstruct the underlying population distribution from a truncated sample, unless the detailed information of the intrinsic correlation is accurately known in advance. In addition, we argue that an intrinsic correlation between luminosity function and formation rate is unlikely eliminated by a misconfigured transformation, and the $g(z)$, derived from a truncated sample with the $\tau$ statistical method, does not necessarily represent its underlying luminosity evolution.
\end{abstract}

\begin{keywords}
{(transients:) gamma-ray bursts - methods: numerical - stars: luminosity function, mass function}
\end{keywords}

%%%%%%%%%%%%%%%%%%%%%%%%%%%%%%%%%%%%%%%%%%%%%%%%%%

%%%%%%%%%%%%%%%%% BODY OF PAPER %%%%%%%%%%%%%%%%%%

\section{Introduction}\label{sec:intro}

For any astronomical source, there are two key properties that characterises the population: (a) their cosmic formation rate, representing the number of sources per unit comoving volume and time as a function of redshift; (b) their luminosity function, which represents the relative fraction of sources in a given luminosity range per unit volume. The statistical problem at hand is the determination of the two properties from flux-limited samples. \citet{1971MNRAS.155...95L} applied a novel method to study the luminosity function and density evolution from a flux-limit quasar sample, which is called Lynden-Bell's $c^{-}$ method. This method is based on the assumption that the luminosity is independent of the redshift. To  overcome this shortcoming, \cite{1992ApJ...399..345E} generalized Lynden-Bell's idea and developed a non-parametric test statistic for independence, which is called non-parametric $\tau$ statistical method. 
These methods have been widely used to estimate the intrinsic luminosity function and cosmic formation rate  of astronomical sources, such as galaxies \citep{1978AJ.....83.1549K,1986MNRAS.221..233P,1986ApJ...307L...1L}, GRBs\citep{2004ApJ...609..935Y,2002ApJ...574..554L,2014ApJ...789...65Y,2006ApJ...642..371K}, and quasars\citep{1999ApJ...518...32M,2011ApJ...743..104S}. 

Before applying Lynden-Bell's $c^{-}$ method, one must first determine whether L and z are correlated or not. Traditionally, if correlated, many authors (e.g., \cite{1999ApJ...511..550L,1999ApJ...518...32M,2002ApJ...574..554L,2004ApJ...609..935Y,2015ApJS..218...13Y}) parameterized it as the transformation, $L^{\prime}=L/g(z)$, where $g(z)=(1+z)^{k}$, the power-law redshift dependence is always adopted to parameterize the luminosity evolution. Once a function $g(z)$ is found, one could remove the correlation and yield an uncorrelated data set $\{L^{\prime},z\}$, then their distributions could be estimated by using Lynden-Bell's $c^{-}$ method.

By using this method to derive the luminosity function and the formation rate of GRBs, \cite{2015ApJS..218...13Y} found that an unexpectedly low-redshift excess in the formation rate of GRBs, compared to the star formation rate (SFR). Whereas, following the same approach, other authors \citep{2016A&A...587A..40P,2017ApJ...850..161T} did not.

More recently, \cite{2021MNRAS.504.4192B} had re-analysed several previous works \citep{2015ApJS..218...13Y,2016A&A...587A..40P,2017ApJ...850..161T,2019MNRAS.488.5823L} and investigated the origin of the evolution of the luminosity/energetics of GRBs with redshift based on the same approach, and found that the effects of the detection threshold have been likely severely underestimated. Then they argued that the observed correlations are artefacts of the individually chosen detection thresholds of the various gamma-ray detectors, and that an inappropriate use of this statistical method could lead to biased scientific discoveries.

So our questions arise: When one applies a truncation function to an intrinsically correlated population, what factors would impact on the distribution of the $\tau$ statistic? In the case, does the transformation decouple the intrinsic correlation between luminosity and redshift of astronomical sources? 

In this paper, we will investigate the issues in detail by performing Monte Carlo simulations. An introduction to Lynden-Bell’s $c^{-}$ method and the nonparametric $\tau$ statistical method is given in Section \ref{sec:method}. To investigate the factors affecting the $\tau$ statistic, in Section \ref{sec:Norm}, we first apply the $\tau$ statistical method to a toy model, where the intrinsic correlations between two random variables are known. Then, in Section \ref{sec:LF}, applying the same approach to the realistic example in an astronomical context, i.e., the luminosity function and the formation rate of GRBs, we explore whether a correlated population distribution could be unbiasedly reconstructed by the transformation.  Finally, in Section \ref{sec:diss}, we present our conclusions and discussions. 

\section{LYNDEN-BELL'S $c^{-}$ METHOD AND NONPARAMETRIC TEST METHOD}\label{sec:method}

Following the description about Lynden-Bell's $c^{-}$ method in \citep{2020sdmm.book.....I} (see Figures of (4.8) and (4.9) in their Book), Here we give a summary description for the test statistic as follows.

Suppose $X$ and $Y$ are the two random variables (RVs), if they are uncorrelated, the bivariate joint density of $(x,y)$ can be represented as 
\begin{equation}\label{eq:uncorelated}
h(x,y) = f(x)g(y).
\end{equation}

Assuming that pairs $(x,y)$ are observable only if they satisfy the truncation function \citep{1992ApJ...399..345E},
\begin{equation}\label{eq:truncation}
y \leq S(x),
\end{equation} 
here $S(x)$ is a monotonic function of $x$. In an astronomical context, $x$ can be considered as redshift, $y$ as absolute magnitude, and the truncation function as magnitude limit, as shown in Figure 4.10 in \cite{2020sdmm.book.....I}.

\begin{figure}
\centering
\includegraphics[scale=0.45]{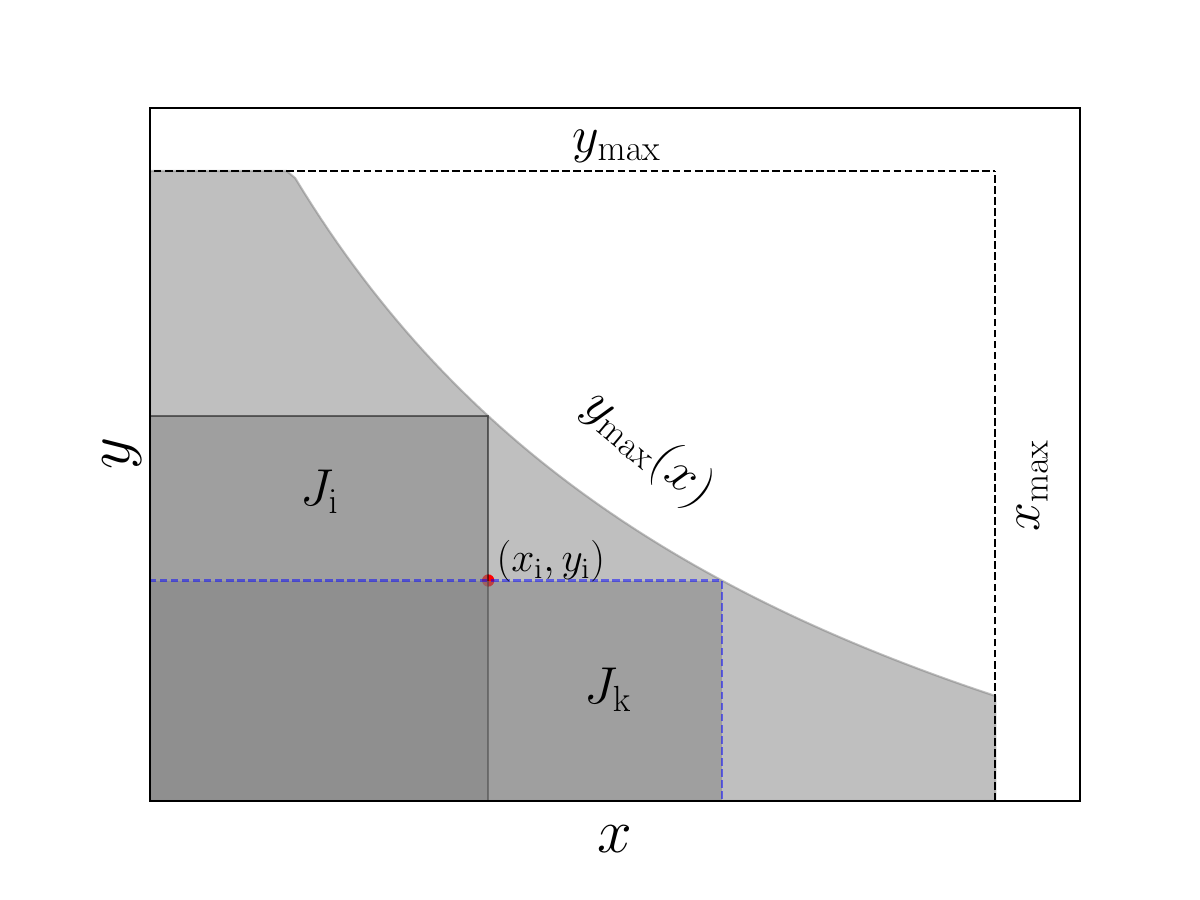}
\caption{Illustration for the definition of a truncated data set, and for the associated subset used by Lynden-Bell $c^{-}$ method. The truncated function is defined by Eq. (\ref{eq:selection}). The sample is limited by $x < x_{\rm max}$ and $y < y_{\rm max}(x)$ (light-shaded area). Associated sets $J_{\rm i}$ and $J_{\rm k}$  are shown by the black solid rectangle and blue dashed rectangle, respectively. Noted this figure is adapted from Fig. (4.8) in \citet{2020sdmm.book.....I}  }
	\label{toy}
\end{figure}

In the following analysis, we will adopt a symmetric function (\cite{2020sdmm.book.....I} (seen in Fig.(\ref{toy})) as 

\begin{equation}\label{eq:selection}
S(x)=\frac{1}{x+t_{\rm low}}-t_{\rm low} ~,
\end{equation}
where $t_{\rm low}$ is a constant.
To test for independence when the data is truncated,  firstly, for \emph{i}th object, one can define an associated set as 
\begin{equation}\label{eq:Ji}
J_{\rm i} = \{ j|x_{j} \leq x_{\rm i}, y_{\rm j} \leq y_{\rm max}(x_{\rm i})  ~, i=1, 2, ..., n \},
\end{equation} 
where $n$ is the size of the observable sample. This is the largest $x$-limited and $y$-limited data subset for \emph{i}th object, with $N_{\rm i}$ elements. This region is shown in Fig. (\ref{toy}) as a black solid rectangle. 

Secondly, sorting the set $J_{\rm i}$ by $y_{\rm j}$, then the number of objects with $y < y_{\rm i}$ in set $J_{\rm i}$ is defined as $R_{\rm j}$, the rank for \emph{i}th object. If X and Y are independent (Null hypothesis $H_{\rm 0}$), then $R_{\rm j}$ is uniformly distributed between 1 to $N_{\rm i}$. The Efron-Petrosian test statistic $\tau$ is then,  
\begin{equation}\label{eq:tau}
\tau \equiv \frac{\sum_{i}(R_{i}-E_{i})}{\sqrt{\sum_{i}V_{i}}} ~,
\end{equation}
\noindent where $E_{\rm i} = (1+N_{\rm i})/2$, $V_{\rm i} = (N_{\rm i} - 1)^2/12$ are the expected mean and the variance of $R_{\rm i}$, respectively. This is a specialized version of Kendell's $\tau$ statistic. The $\tau$ statistic has mean 0 and variance 1 under $H_{\rm 0}$. As pointed out by \cite{1992ApJ...399..345E}, the $\tau$ statistic will approximately follow as a standard normal distribution, $N(0,1)$, even for $n$ as small as 10 under $H_{\rm 0}$.

Following classical statistical inference, one can accept $H_{\rm 0}$ if $\tau\leq 1.645$, and reject $H_{\rm 0}$ otherwise. The rejection probability of the test for independence would be approximately 0.10. Once accepting $H_0$, the cumulative distributions for the two random variables are defined as \citep{2020sdmm.book.....I}
\begin{equation}\label{Fx}
F_{\rm X}(x) =  \int_{-\infty}^{x} f(x') \, dx' 
\end{equation}
and 
\begin{equation}\label{Fy}
G_{\rm Y}(y) =  \int_{-\infty}^{y} g(y') \, dy'.
\end{equation}

Then, 
\begin{equation}\label{Fx1}
F_{\rm X}(x_{i}) = F_{\rm X}(x_{1}) \prod\limits_{k=2}^{i}(1+\frac{1}{N_k}) ~,
\end{equation}
where it is assumed that $x_{\rm i}$ are sorted ($x_{\rm 1}  \leq x_k \leq  x_{\rm n}$). Analogously, if $M_{\rm k}$ is the number of objects in an associated set defined as $J_{\rm k} = \{ j|y_{\rm j} \leq y_{\rm k}, x_{\rm j} \leq x_{\rm max}(y_{\rm k}) \}$. This region is also shown in Fig. (\ref{toy}) as a blue dashed rectangle. Then 
\begin{equation}\label{Fy1}
G_{\rm Y}(y_{j}) = G_{\rm Y}(y_{1}) \prod\limits_{k=2}^{j}(1+\frac{1}{M_k}) ~,
\end{equation}
where it is also assumed that $y_{\rm j}$ are sorted ($y_{\rm 1}  \leq y_{\rm k} \leq  y_{\rm n}$).

\section{Case of The Bivariate Normal Distribution}\label{sec:Norm}

A bivariate distribution is a statistical method used to show the probability of occurrence of two random variables. In this section, we first use Monte Carlo simulation to test the $\tau$ statistic based on a truncated bivariate normal distribution when the relations between two random variables are known (similar to the toy model in \cite{2020sdmm.book.....I}), and next, we further investigate whether one can apply the $\tau$ statistical method to decouple a correlation between the physical quantities of a source, such as luminosity and redshift, generated by effects of evolution and the truncation or bias introduced by the flux limit.

\begin{figure}
	\centering
	\includegraphics[scale=0.45]{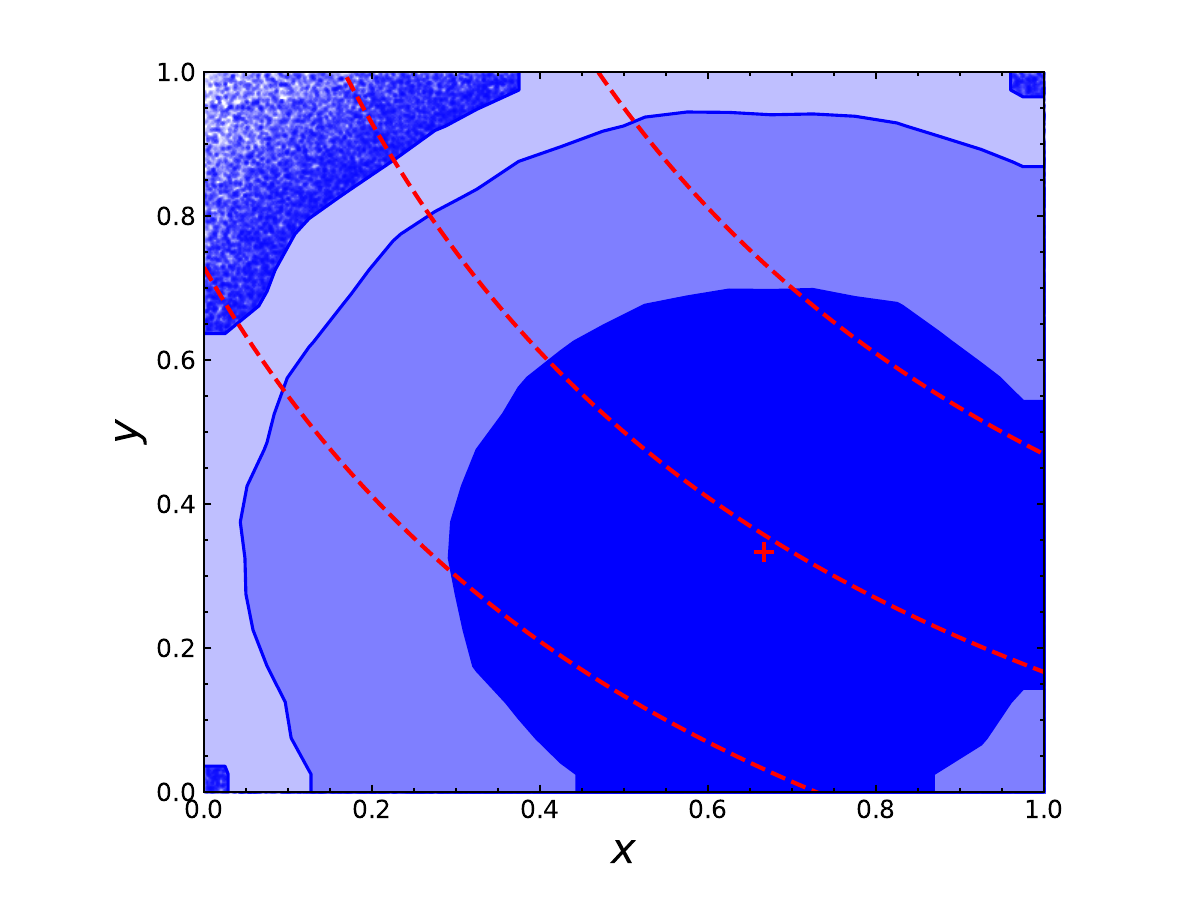}
	\includegraphics[scale=0.45]{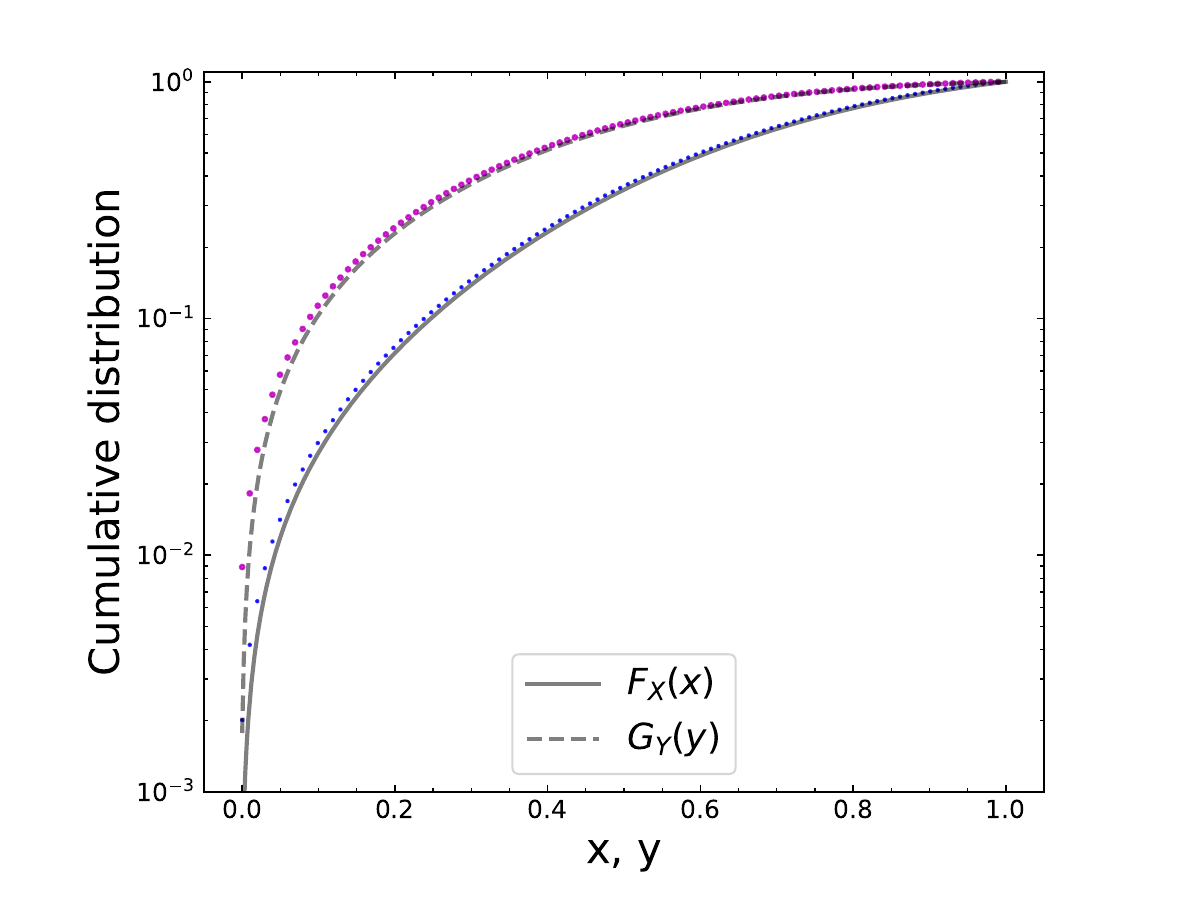}
	\caption{Illustration of a realization of the truncated bivariate normal distribution with the parameters of ($\mu_{x},\mu_{y})=(0.67,0.33)$, ($\sigma_{x},\sigma_{y})=(0.33,0.33)$, and $r=0$. Upper panel: the corner of a sample from the parent distribution with $n=10^6$, in which the contours enclose the regions with different colors that contain 68\%, 95\% and 99\%  of the probability, the red dashed lines from upper to lower represent the truncated function defined in Eq. (\ref{eq:selection}) with $t_{\rm low}=0.3, 0.5, 0.7$, respectively, and the red plus marks the center of the PDF. Lower panel: comparison of the one-dimensional cumulative distributions (solid points) of the sample to their corresponding population distributions.}
	\label{Lyndentoy}
\end{figure}

RVs $X$ and $Y$ have a joint normal distribution, $p(X,Y)\sim N(\mu, \Sigma)$. First step,  we investigate this issue based on the assumption that there is none  correlation between RVs $X$ and $Y$. We define a joint probability density function (PDF) from  a $2D$ truncated normal distribution based on the \textbf{truncMVN} python package\footnote{https://github.com/zachjennings/truncMVN}, with the parameters of ($\mu_{X},\mu_{Y})$=(0.67,0.33), ($\sigma_{X},\sigma_{Y})$=(0.33,0.33), and $\Sigma_{X,Y}=0$, and $X$ and $Y$ are truncated in the range of [0,1]. Then we generate a random sample with the size of $n=10^6$ objects from the population distribution with the \textbf{emcee} sampler\footnote{https://emcee.readthedocs.io/en/stable/index.html}. Shown in Fig. (\ref{Lyndentoy}) are the corner and the resultant one-cumulative distributions of the sample. We obtain the sample Pearson correlation coefficient, $r\simeq0$. To study the effect of selection (or truncated) function on the $\tau$ statistic, we apply three different selection functions with $t_{\rm low}=0.7, 0.5, 0.3$ to the sample, respectively, and then obtain three truncated data sets. As seen in the upper panel of Fig. (\ref{Lyndentoy}), the larger the $t_{\rm low}$, the smaller the proportion of truncated samples in the population.

\begin{figure}
	\centering
	\includegraphics[scale=0.44]{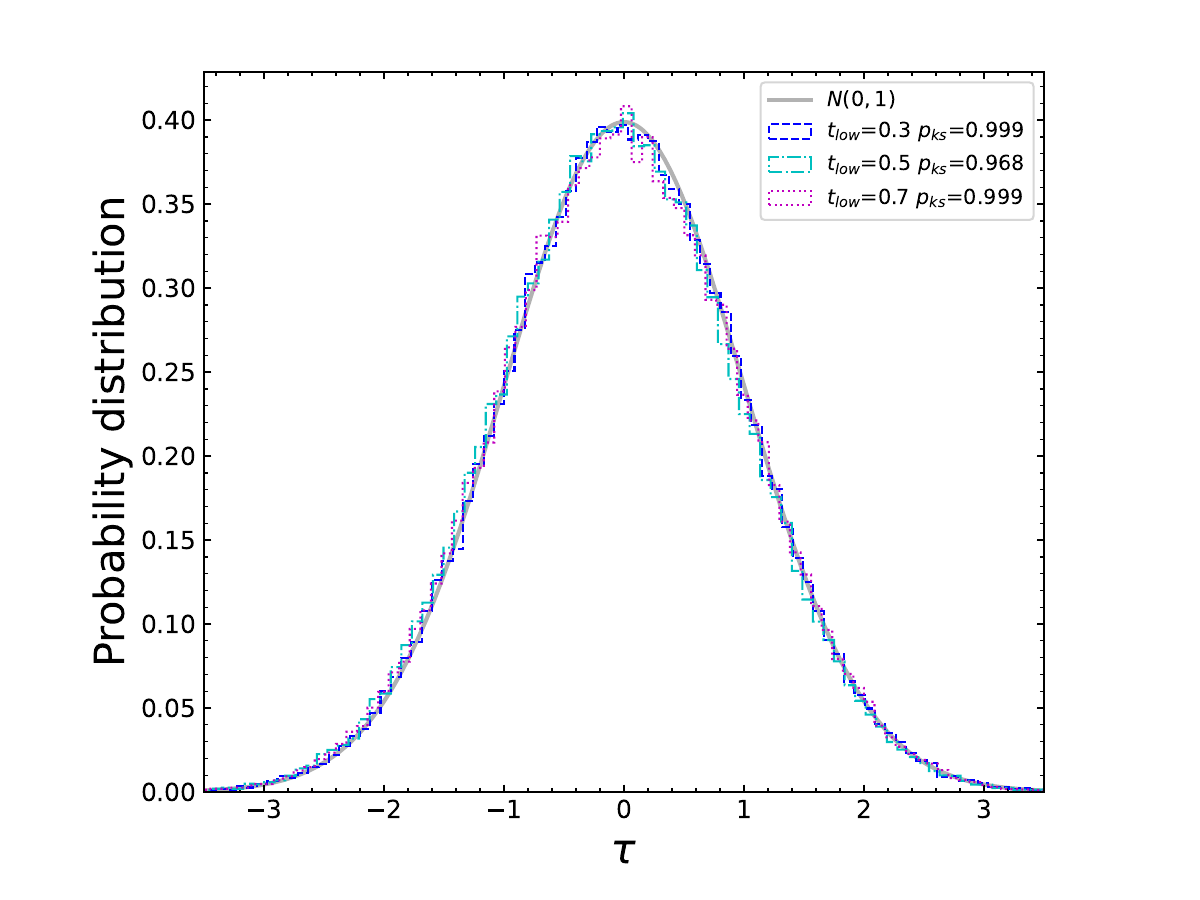}
	\caption{Comparison of the distributions of the $\tau$ statistic derived from the three truncated data (See the text for details) to the standard normal distribution (solid line). The chance probabilities of $KS$-tests are also presented in their corresponding legends.}
	\label{taus}
\end{figure}

For every truncated data set, we create $10^5$ pseudo samples and each sample contains $10^2$ observable objects\footnote{Our analysis shows that the distribution of the $\tau$ statistic is less sensitive to the size of observable sample, as pointed out in \cite{1992ApJ...399..345E} that the $\tau$ has a short-tailed distribution if RVs $X$ and $Y$ are uncorrelated. So here we also use a sample of $10^2$ objects for our analysis.}. Then we calculate the distribution of the $\tau$ statistic based on Eq.(\ref{eq:tau}) for these pseudo samples, and compare it to a standard normal distribution by using a KS-test. The results are shown in Fig. (\ref{taus}). The chance probabilities of the three tests are 0.999, 0.968 and 0.999, respectively, which indicates that the $\tau$ statistic follows well a standard normal distribution for all the three cases. Thus, with Monte Carlo simulations, for the first time, we confirm the conclusions proposed by \cite{1992ApJ...399..345E}, i.e., the $\tau$ statistic always follow a standard normal distribution for any truncated data as long as $H_0$ does hold.

\begin{figure}
	\centering
	\includegraphics[scale=0.44]{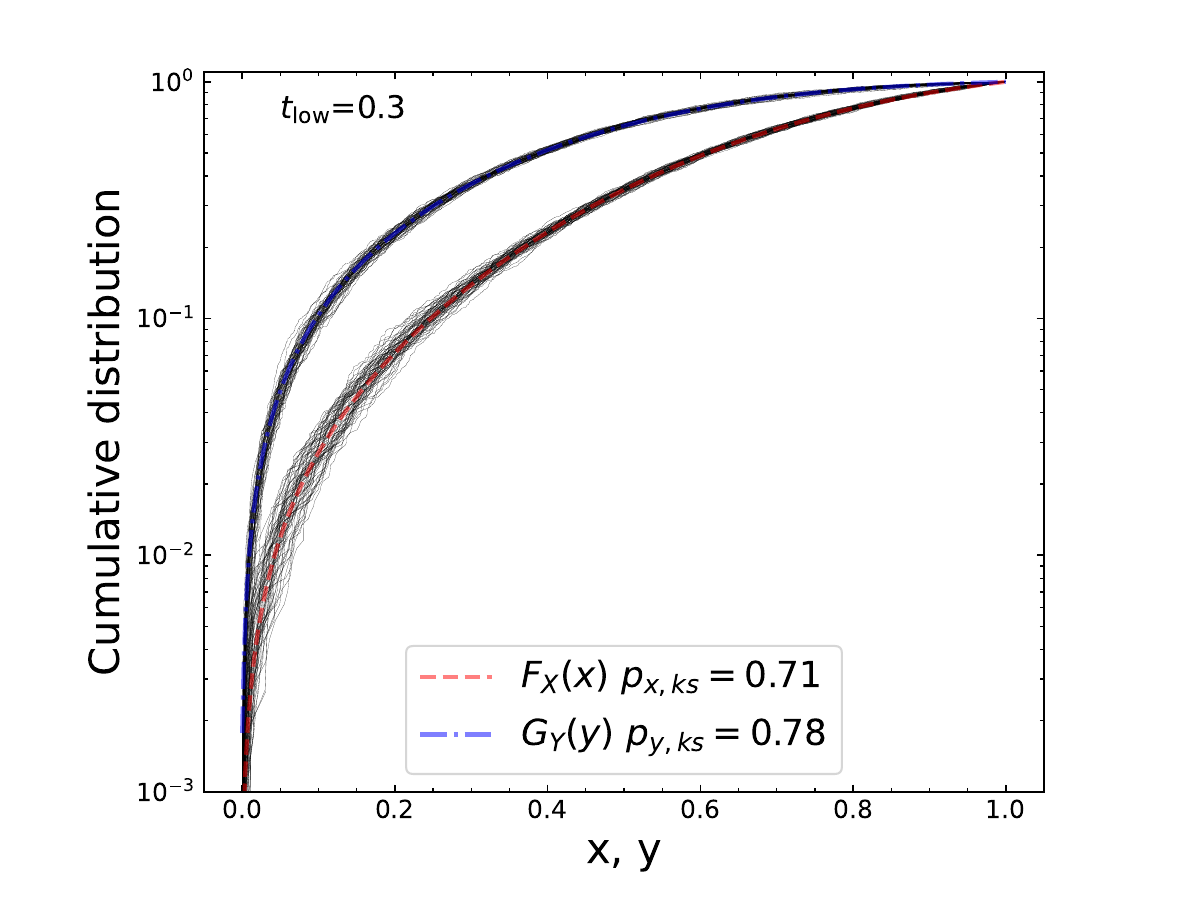}
	\includegraphics[scale=0.44]{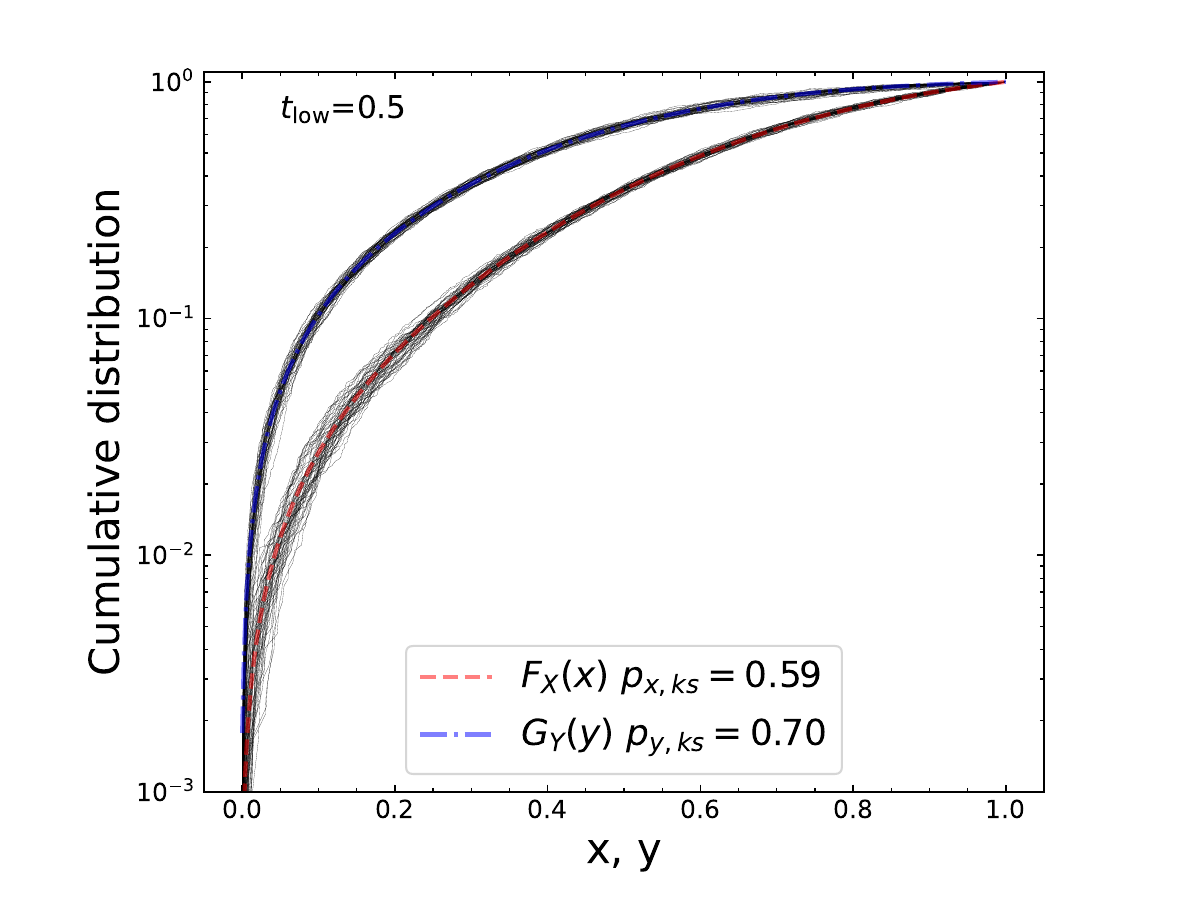}
	\includegraphics[scale=0.44]{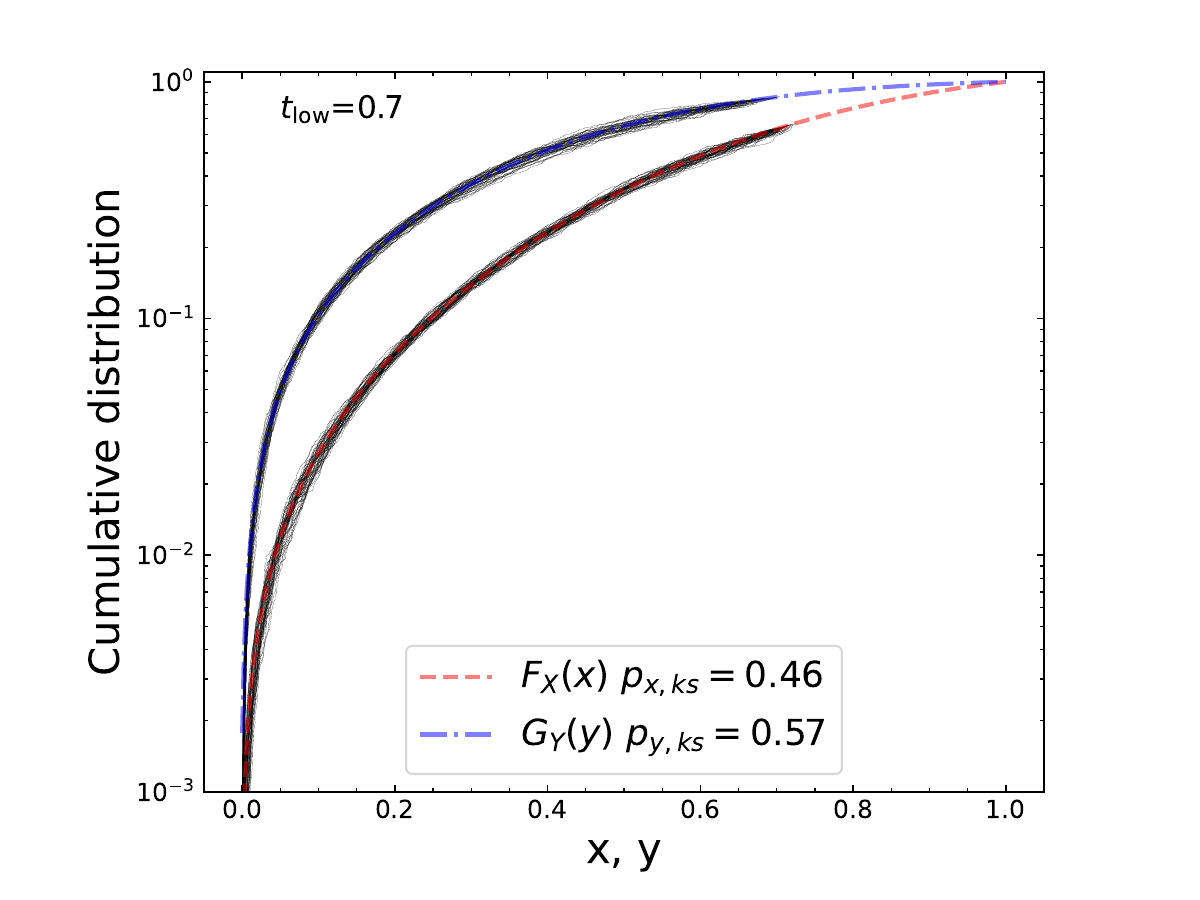}
	\caption{Demonstrations of the performances of the reconstructed distributions from different truncated data, compared to their corresponding population distributions. In each panel, the red dashed lines and the blue dash-dotted lines represent the population distributions of $X$ and $Y$, respectively. The truncated functions with different values of $t_{\rm low}$ are also marked in their corresponding panels, respectively.  The chance probabilities of $KS$-tests between the mean distributions of those truncated data and population distributions are also presented in their corresponding legends. For every random variable, $X$ and $Y$, in each panel, we only plot 50 samples (gray solid lines) chosed randomly from all truncated samples.}
	\label{cumulative}
\end{figure}

At the same time, with these pseudo truncated samples, we also derive the one-dimensional cumulative distributions for the two random variables based on Eqs. of (\ref{Fx1}) and (\ref{Fy1}), and compare them to their corresponding population distributions by the KS-test, respectively. The comparison results are shown in Fig. (\ref{cumulative}), which indicates that their corresponding population distributions could be unbiasedly recovered. Note that here we normalize the sample distribution to its corresponding population distribution for comparison. Further investigations show that the reconstructed cumulative distributions are not sensitive to the truncated sample size we adopted.  

In conclusion, the $\tau$ statistic is indeed a robust test statistic for independence of the truncated data. Once one accepts $H_0$ (i.e., $X$ and $Y$ are truly independent), their population distribution could be unbiasedly reconstructed from truncated data with Eqs. of (\ref{Fx1}) and (\ref{Fy1}), irrespective of adopted selection functions. The fact shows that an observed correlation does unlikely come from a truncation effect as long as the $H_0$ does hold.

Now we wonder whether the $\tau$ statistic still obeys a standard normal distribution when the two random variables, $X$ and $Y$, are correlated. In that case, whether one would still unbiasedly recover the underlying population distribution from truncated data based on Eqs. of (\ref{Fx1}) and (\ref{Fy1}) or not? 

With the \textbf{truncMVN} python package and following the same sampling method above, we can also create some pseudo samples from the population distributions with the different values of $\mu$ and $\Sigma_{X,Y}$, as adopted in Fig. (\ref{Lyndentoy}). As pointed out by some authors \citep{Raymond,2020arXiv200913488G}, the moment of the truncated variable depends on the mean and covariance matrix of a bivariate population distribution, and one can find the explicit expression for low order moments of the truncated multivariate normal distribution in their papers. In this paper, we calculate the Pearson correlation coefficient ($r$) of the pseudo sample  with numpy python library\footnote{https://numpy.org/}. Finally, by sampling from different populations with different values of $\mu$ and $\Sigma_{X,Y}$, we obtain four samples with Pearson's $r$=0.15, 0.32, 0.50, and 0.73, respectively. 

\begin{figure}
	\centering
	\includegraphics[scale=0.450]{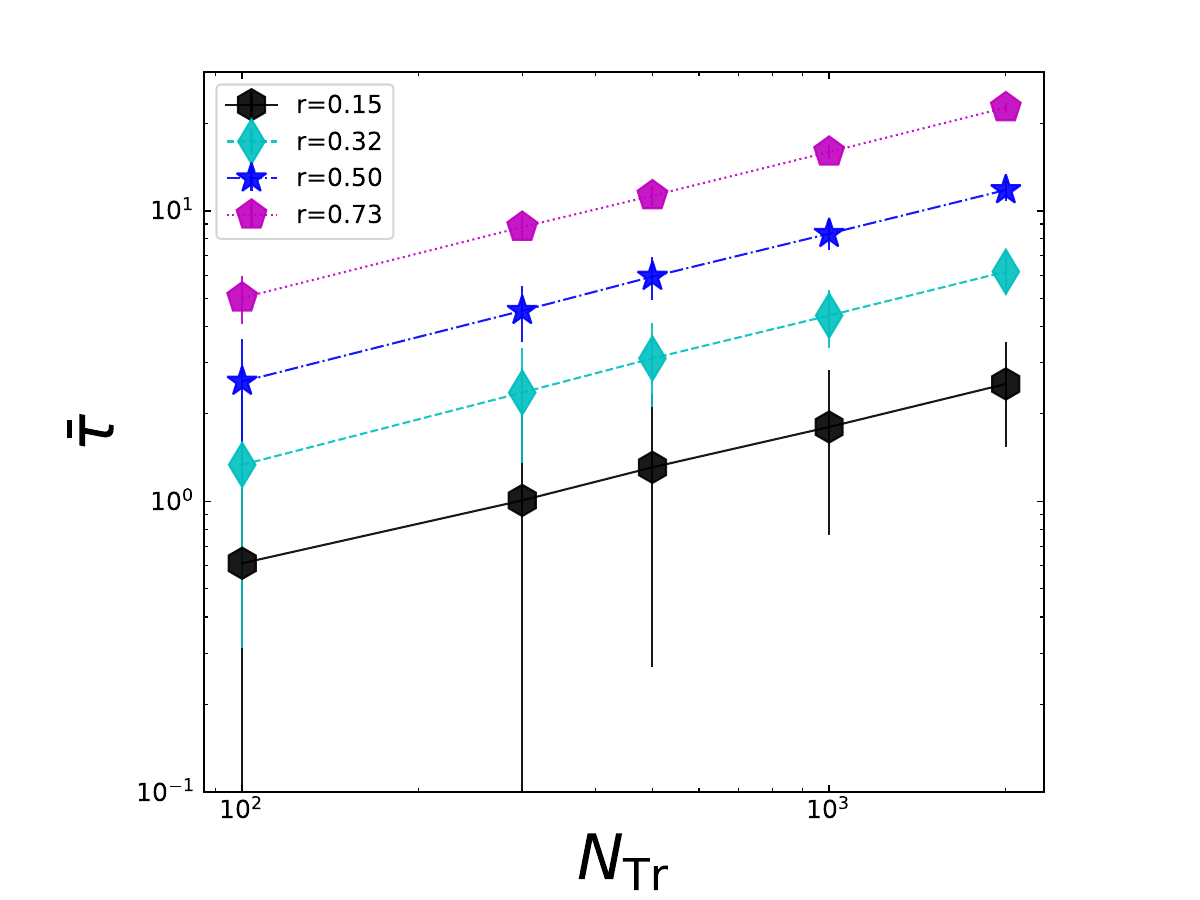}
	\includegraphics[scale=0.450]{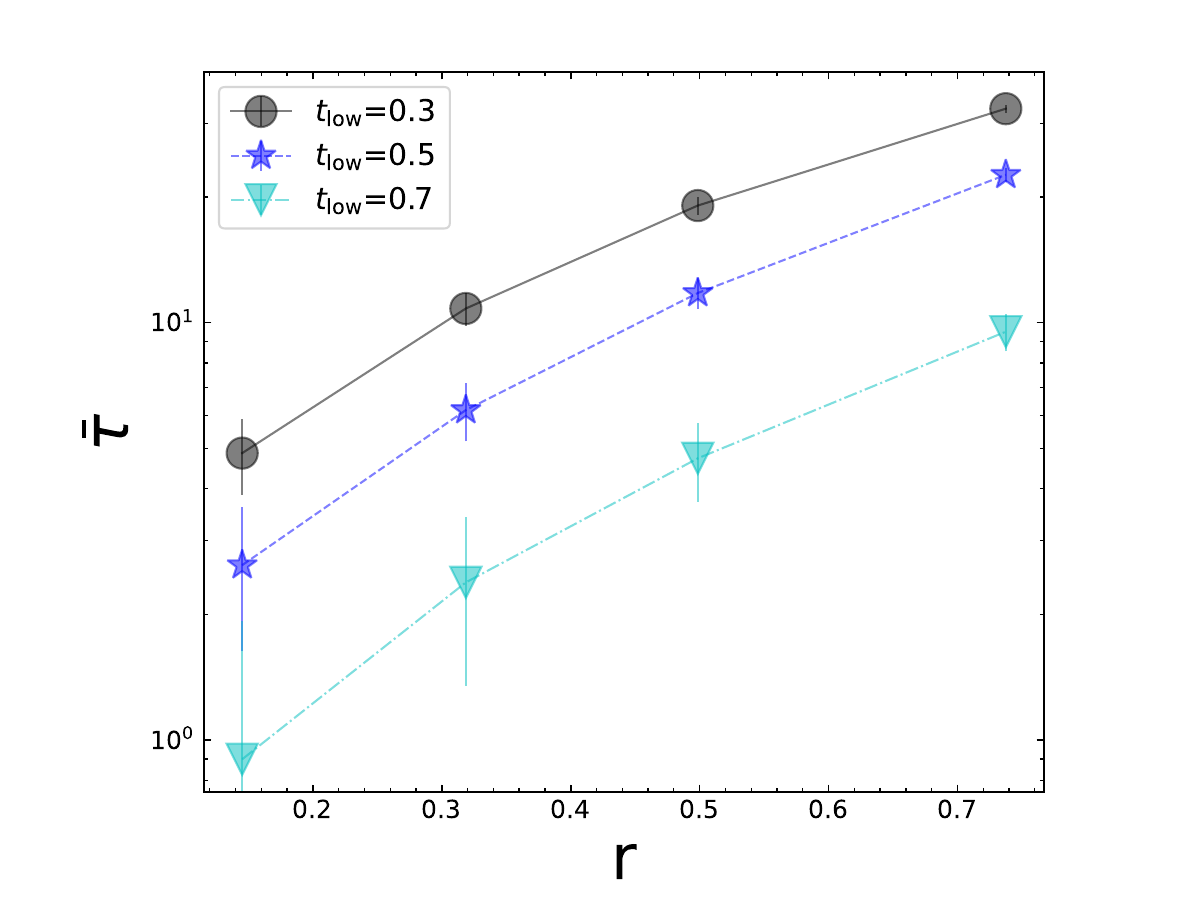}
	\caption{Demonstrations of the factors impacting on the $\tau$ statistic. Upper panel:  the $\bar{\tau}$ is a function of the observable sample size ($N_{\rm Tr}$) coming from different populations with different values of $r$ (as marked in the legend), but with the same selection function with $t_{\rm low}=0.5$. Lower panel: the $\bar{\tau}$ is a function of the Pearson’s coefficient $r$ of population, and the observable samples with the same of $N_{\rm Tr}=2\times10^3$ come from the selection functions with different values of $t_{\rm low}$ (as marked in the legend). For all cases, the value of $s_{\rm \tau}$ is marked by the error bar of every data point.  }
	\label{tau-N}
\end{figure}

To answer the questions mentioned above, we also apply the same three truncation functions above to the four samples, respectively, and obtain three truncated data sets for every sample.  Then, with the same methods as done in Figs. of (\ref{taus}) and (\ref{cumulative}), we calculate the distributions of the $\tau$ statistic and their corresponding one-dimensional cumulative distributions for RVs $X$ and $Y$, respectively. Meanwhile, the influence of observable sample size on the $\tau$ statistic is also explored. Finally, the results of these analyses are shown in Fig. (\ref{tau-N}).

Evidently, unlike the case of an uncorrelated bivariate distribution, we can draw the following conclusions from Fig. (\ref{tau-N}):

In all cases we have explored, the $\tau$ statistic no longer follow a standard normal distribution, but a normal distribution, defined as $N(\bar{\tau},s^2_{\tau})$ (where $\bar{\tau}$ and $s_{\tau}$ are the average and standard variance of the distribution, respectively). The $\bar{\tau}$ changes with the Pearson's $r$ of population, observable sample size ($N_{\rm Tr}$), as well as different selection functions. But the $s_{\tau}$ is less sensitive to them, with $s_{\tau}\sim 1$ in all cases (The $s_{\tau}$  is shown as the error bar of the data point in the figure), which means that the origin of an observed correlation, measured by the $\tau$ statistical method, is a complex combination of the three factors.

Further investigations show that, in that case, the underlying population distribution can not unbiasedly reconstructed  from the truncated data based on Eqs. of (8) and (9).


\section{Case of the luminosity function and the formation rate of GRBs} \label{sec:LF}
Now, let's turn our attention to the realistic example in an astronomical context. It is often the case in the analysis of astronomical data that one is faced with reconstructing the joint bivariate distribution from truncated data. And the luminosity function and cosmic formation rate of GRBs (or other astronomical object) $(L,z)$ is one such set of bivariate data. For simplicity, it is often assumed that such a bivariate distribution is separable in the following form,
\begin{equation}   
\label{eq:psi}
\Psi(z,L) = \rho(z)\phi(L),
\end{equation}
\noindent where $\rho(z)$ is the GRB Formation Rate and $\phi(L)$ is GRB's LF. 

The GRB formation rate is usually assumed to trace the cosmic star formation rate (SFR). Here, we assume that the rate is purely proportional to the SFR, and parameterize it as \citep{2006ApJ...651..142H,2008MNRAS.388.1487L}
\begin{equation}
\rho(z)=\frac{0.0157+0.118z}{1+(z/3.23)^{4.66}}\;.
\end{equation}

For the GRB's LF, we adopt the Schechter function \citep{1976ApJ...203..297S} as follows
\begin{equation}\label{eq:Schechter}
\phi(L) = \left( \frac{\phi^*}{L^*} \right) \left( \frac{L}{L^*} \right)^\alpha \exp{\left( -\frac{L}{L^*} \right)},
\end{equation}
where $\alpha$ represents the power-law parameter for the faint-end and $L^*$ is the characteristic luminosity, while $\phi*$ serves as the normalisation constant. Due to the typically large span of the luminosities, here we use logarithmic units in the Schechter function, written as follows,
\begin{equation}\label{Schechter}
\phi(\log L) = \ln 10 \phi^* 10^{(\alpha + 1)(\log L - \log L^*)} \exp{\left( -10^{\log L - \log L^*} \right)}
~. 
\end{equation} 
In the following analysis, we adopt arbitrary parameter values:  $\alpha =-0.48$ and $\log (L^{*} / erg s^{-1})=51.2$. 

\begin{figure}
	\centering
	\includegraphics[scale=0.48]{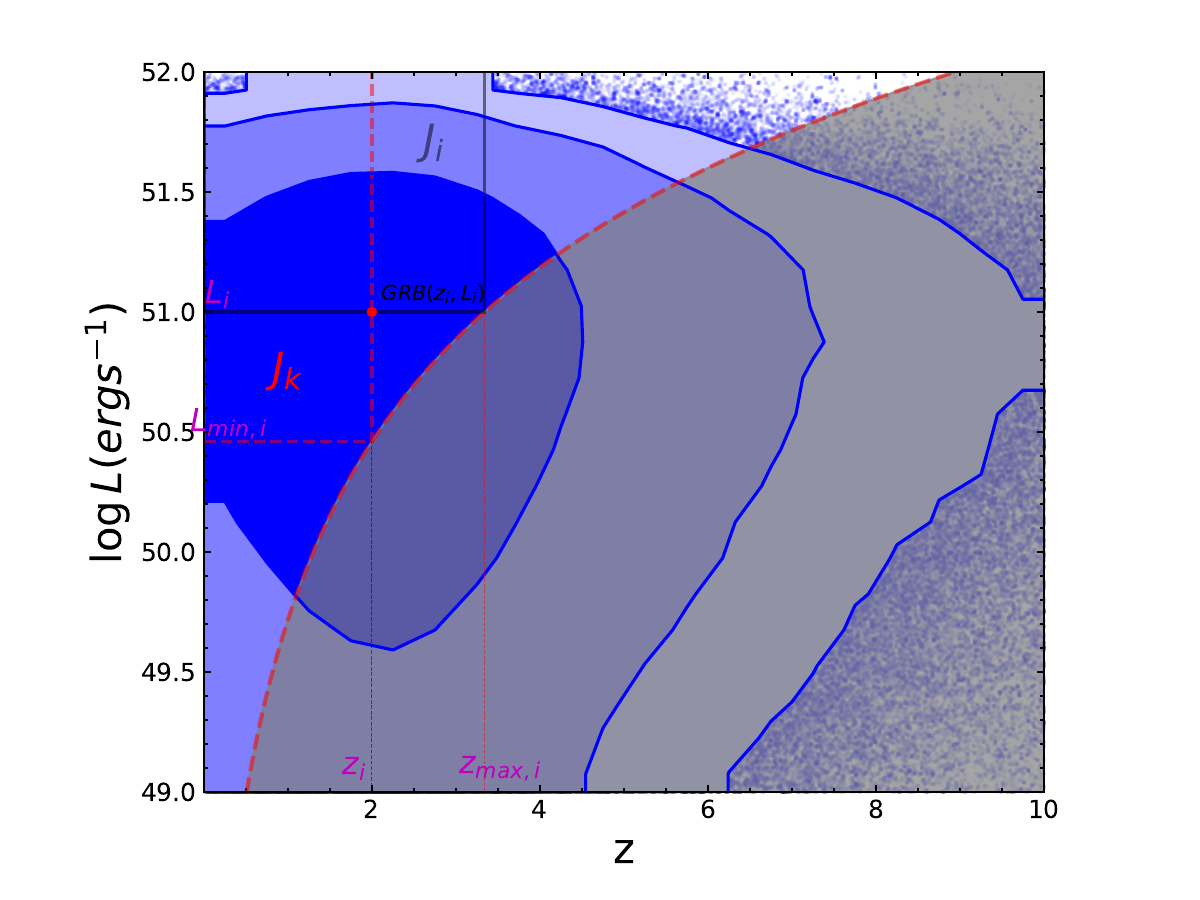}
	\includegraphics[scale=0.48]{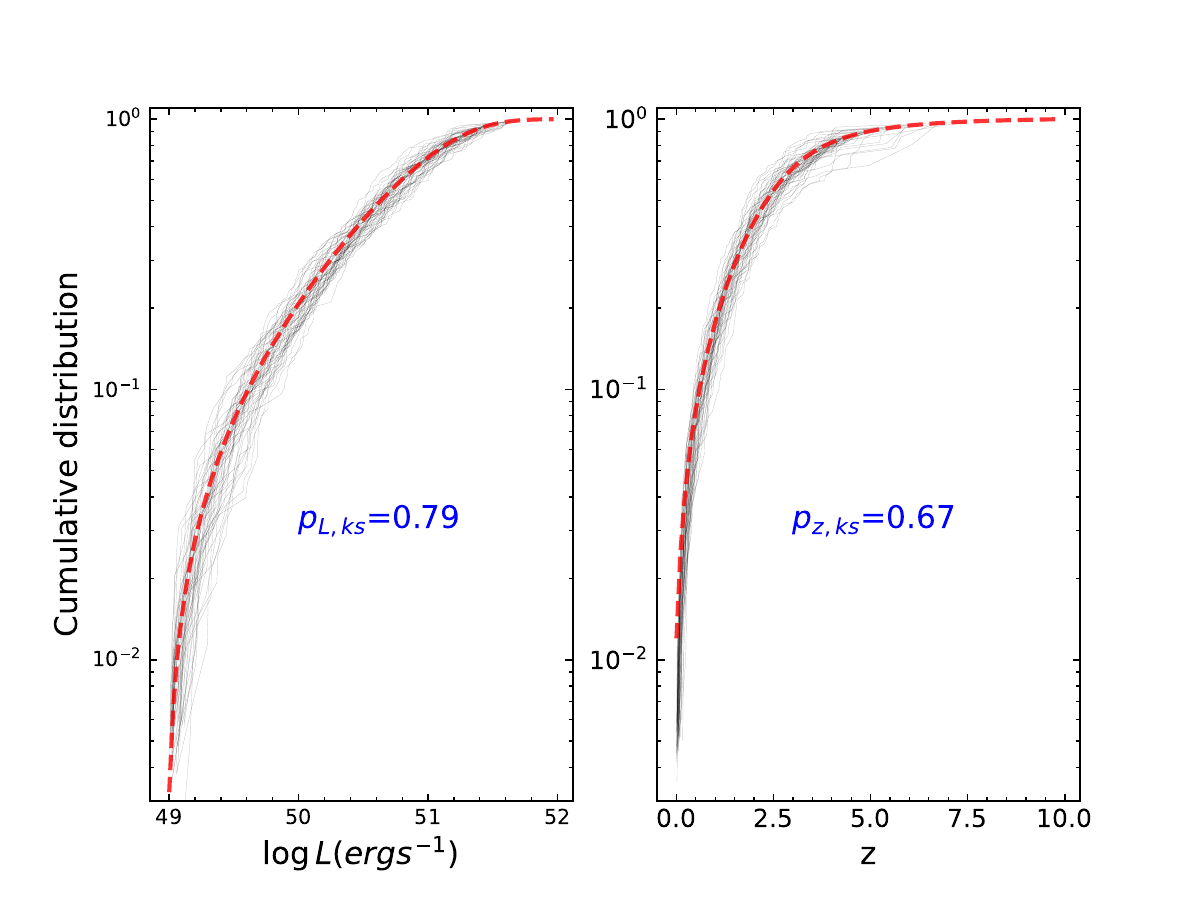}
	\caption{Upper panel: the same as Fig. (\ref{Lyndentoy}), but for the luminosity function. The red dashed line represents the truncation due to flux limit at $F_{\rm lim}=1\times10^{-8} cm^{-2} s^{-1}$, and the data below the red line (gray area) are unobservable. Associated sets $J_{\rm i}$ and $J_{\rm k}$ for \emph{i}th GRB($z_{\rm i}$, $L_{\rm i}$), marked by the red solid circle, are shown by the black solid rectangle and red dashed rectangle, respectively. Lower panel: the reconstructed luminosity and redshift distributions from the truncated sample (bright area in upper panel), are compared to their corresponding population distributions (the red dashed lines), respectively. For the reconstructed distributions, we only plot 50 samples (gray solid lines) chosed randomly from all simulated samples.}
	\label{Fig:LF}
\end{figure}

In the luminosity-redshift plane, the truncated function is the luminosity limit (The red dashed line in Fig.\ref{Fig:LF}), given as  
\begin{equation}
L_{\rm limit}=4\pi d^{2}_{L}(z) F_{\rm min} ~,
\end{equation} \label{flux-limited}
where $d_{\rm L}(z)$ and $F_{\rm min}$ are the luminosity distance at redshift $z$ and flux-limit, respectively. Associated sets $J_{\rm i}$ and $J_{\rm k}$ for \emph{i}th GRB($z_{\rm i}$, $L_{\rm i}$) can be defined as, 
\begin{equation}
J_{i} = {j: L_{j}\geq L_{i}~, z_{j} \leq z_{\rm max, i}} ~, i = 1, 2, ..., n ~~,
\end{equation}
and 
\begin{equation}
J_{k} = {j: L_{j}\geq L_{\rm min, i}~, z_{j} \leq z_{ i}} ~, i = 1, 2, ..., n ~~,
\end{equation}
respectively (seen Fig.\ref{Fig:LF}).

With the same method done in sec. (\ref{sec:Norm}), we also draw a pseudo sample with $n=10^6$ from the joint probability function (Equation \ref{eq:psi}) with the \textbf{emcee} sampler. With  Eqs. of (\ref{Fx1}) and (\ref{Fy1}), we could also reconstruct their population distributions well from the observable data of the pseudo sample truncated by the flux limit at $1\times10^{-8} cm^{-2} s^{-1}$. The results are shown in Fig. (\ref{Fig:LF}). Further investigations show that, when different values of the flux limit are adopted, the population distributions could always be unbiasedly recovered from their corresponding truncated data, indicating that the non-parametric $\tau$ statistical method can be applied to unbiasedly recover the underlying population distributions from truncated data regardless of adopted detection thresholds. The same conclusion could be arrived in the contex of an uncorrelated bivariate normal distribution.

Unfortunately, here's the fact that the luminosities of GRBs are strongly correlated with their redshifts is a common feature \citep{2010MNRAS.406..558Q,2016ApJ...820...66D,2015ApJS..218...13Y,2015ApJ...806...44P,2016A&A...587A..40P, 2019MNRAS.488.5823L}.  In this case, one can not directly apply Lynden-Bell's $c^{-}$ method to reconstruct their underlying parent distribution. 

The popular method to eliminate the correlation (e.g.,\cite{1999ApJ...511..550L,1999ApJ...518...32M,2002ApJ...574..554L,2004ApJ...609..935Y,2015ApJS..218...13Y}) is to parameterize it as the luminosity evolution through the transformation, $L=L_{z}/g(z)$, where $g(z)=(1+z)^{k}$ parameterizes the luminosity evolution. Then one could extract the luminosity evolution by varying $k$ until $\tau=0$. Once the function $g(z)$ is found, one could reconstruct their underlying parent luminosity and redshift distributions from this uncorrelated data set, $\{L,z\}$.    

\begin{figure}
	\centering
	\includegraphics[scale=0.45]{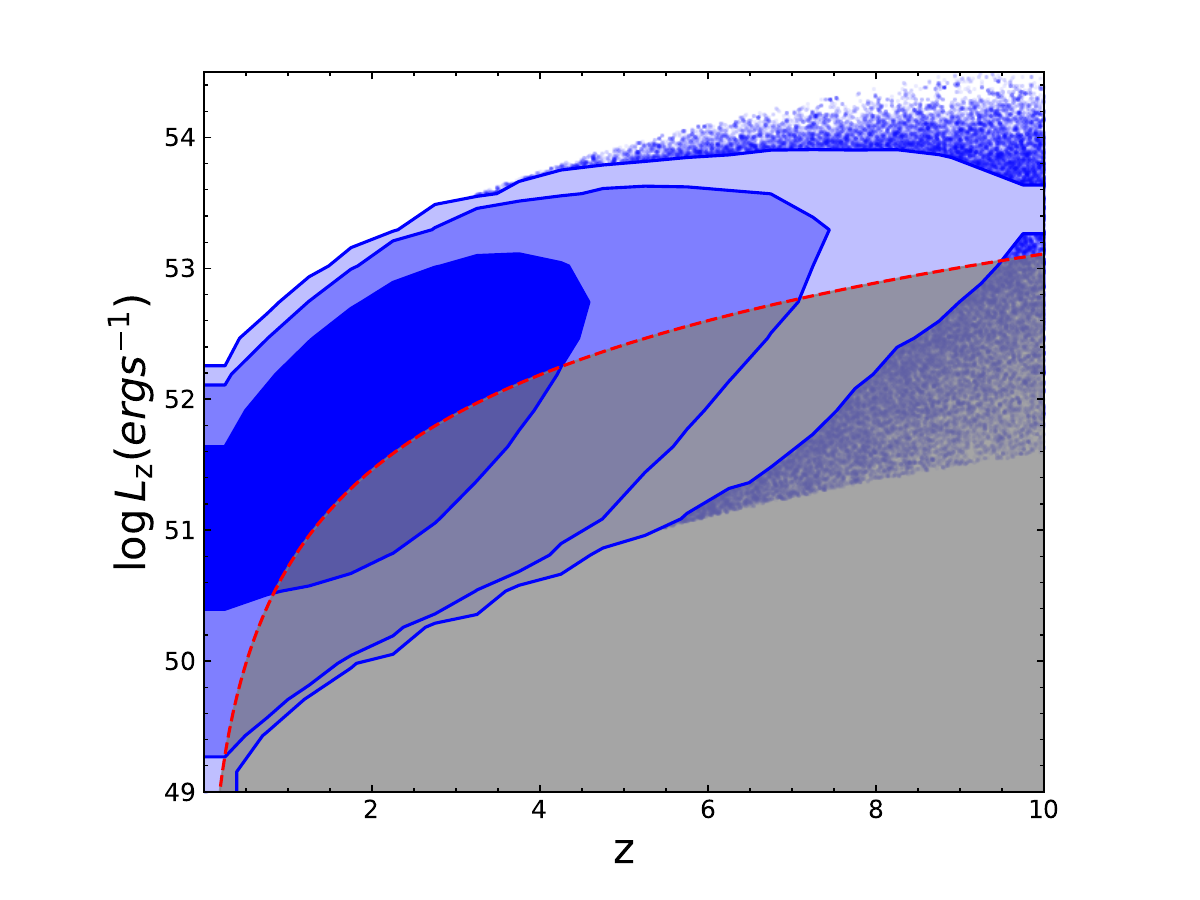}
	\includegraphics[scale=0.45]{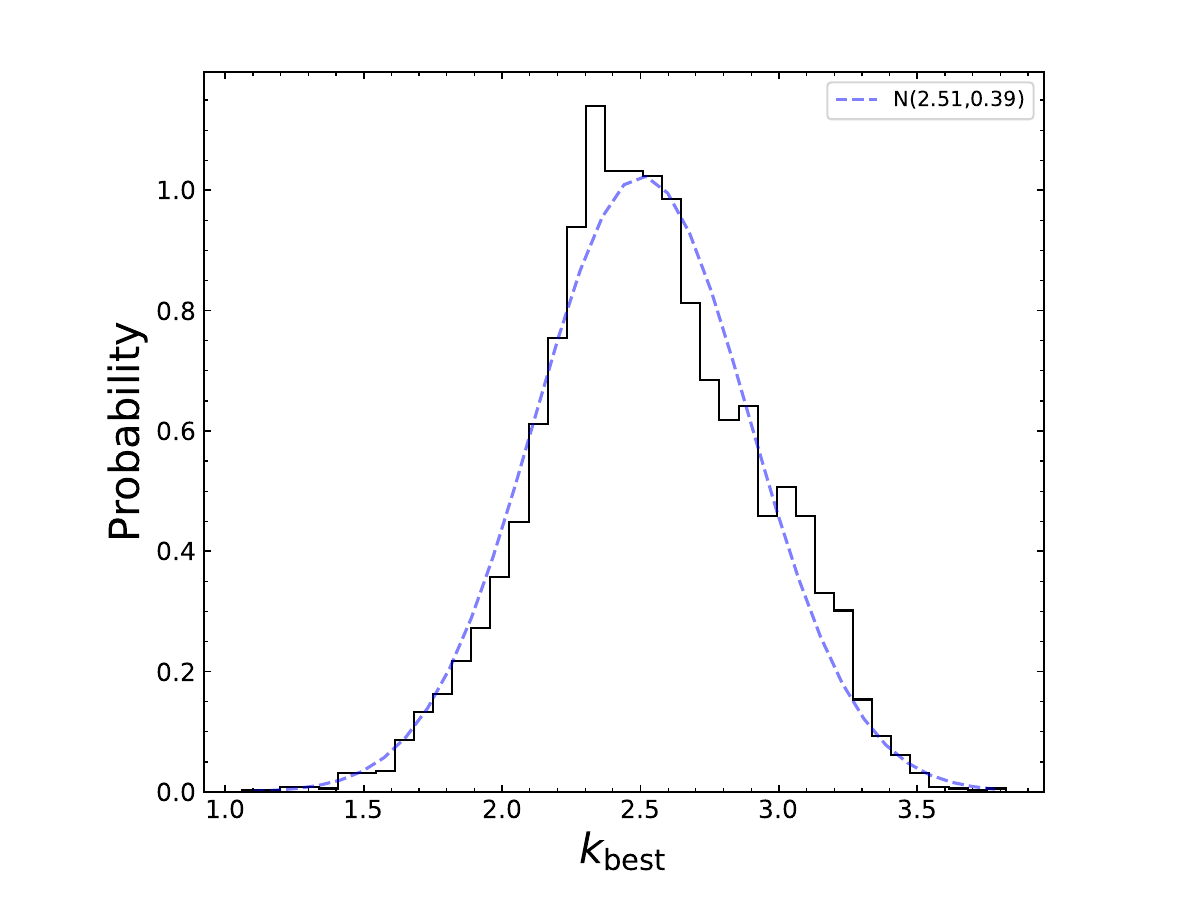}
	\caption{Upper panel: the same as Fig. (\ref{Fig:LF}), but for the intrinsic luminosity evolving with redshift by $L_{z}=L(1+z)^{k}$, where k=2.5. The red dashed line represents the flux limit at $1\times10^{-7} cm^{-2} s^{-1}$. Lower panel: The distribution of the $k_{\rm best}$ at $\tau=0$. The dashed line represents the best fit to a Gaussian function.}
	\label{Fig:LFz}
\end{figure}

Now we verify the correctness of this approach by numerical simulations.  We firstly produce a pseudo correlated sample from the sample shown in the upper panel of Fig. (\ref{Fig:LF}) by the transformation, $L_{\rm z}=Lg_{\rm int}(z)$, where $g_{\rm int}(z)=(1+z)^{k}$, and $k=2.5$ is taken, which means that the information of intrinsic luminosity evolution is known accurately. The corner of the pseudo correlated sample is shown in the upper panel of Fig. (\ref{Fig:LFz})). 

Next, we produce a truncated data set by the flux limit at $1\times10^{-7} cm^{-2} s^{-1}$ (The data above the red line in the upper panel of Fig. (\ref{Fig:LFz}) are observable). Based on the truncated data, we create $10^5$ pseudo samples, and each pseudo sample contains $10^2$ observable GRBs, the same as done in section \ref{sec:Norm}. For every observable sample, we make the reverse transformation of the intrinsic luminosity evolution  $L=L_{z}/g_{\rm int}(z)$, and calculate the best $k$ at $\tau=0$, defining it as $k_{\rm best}$. Next, with Eqs. of (\ref{Fx1}) and (\ref{Fy1}), we could calculate their corresponding luminosity and redshift distributions from these uncorrelated data sets   $\{L_{\rm i},z_{\rm i} \}$. We fit a Gaussian function to the distribution of the $k_{\rm best}$ , giving $\bar{k}_{\rm best}=2.51\pm0.39$. The result is also shown in the lower panel of Fig. (\ref{Fig:LFz}). As the reconstructed luminosity and redshift distributions are similar to those shown in the lower panel of Fig. (\ref{Fig:LF}), we do not plot them repeately. Further investigations show that, the distribution of the $k_{\rm best}$ and the reconstructed luminosity and redshift distribution are less sensitive to both the adopted detector threshold and observable sample size. These results show that, if the detailed information of the luminosity evolution is accurately known, we could remove the effect of the evolution by making the reverse transformation of the intrinsic luminosity evolution, then unbiasedly reconstruct their underlying parent distribution by Lynden-Bell's $c^{-}$ method. Some authors \citep{2015ApJS..218...13Y,2016A&A...587A..40P} came to similar conclusion.  

\begin{figure}
	\centering
	\includegraphics[scale=0.45]{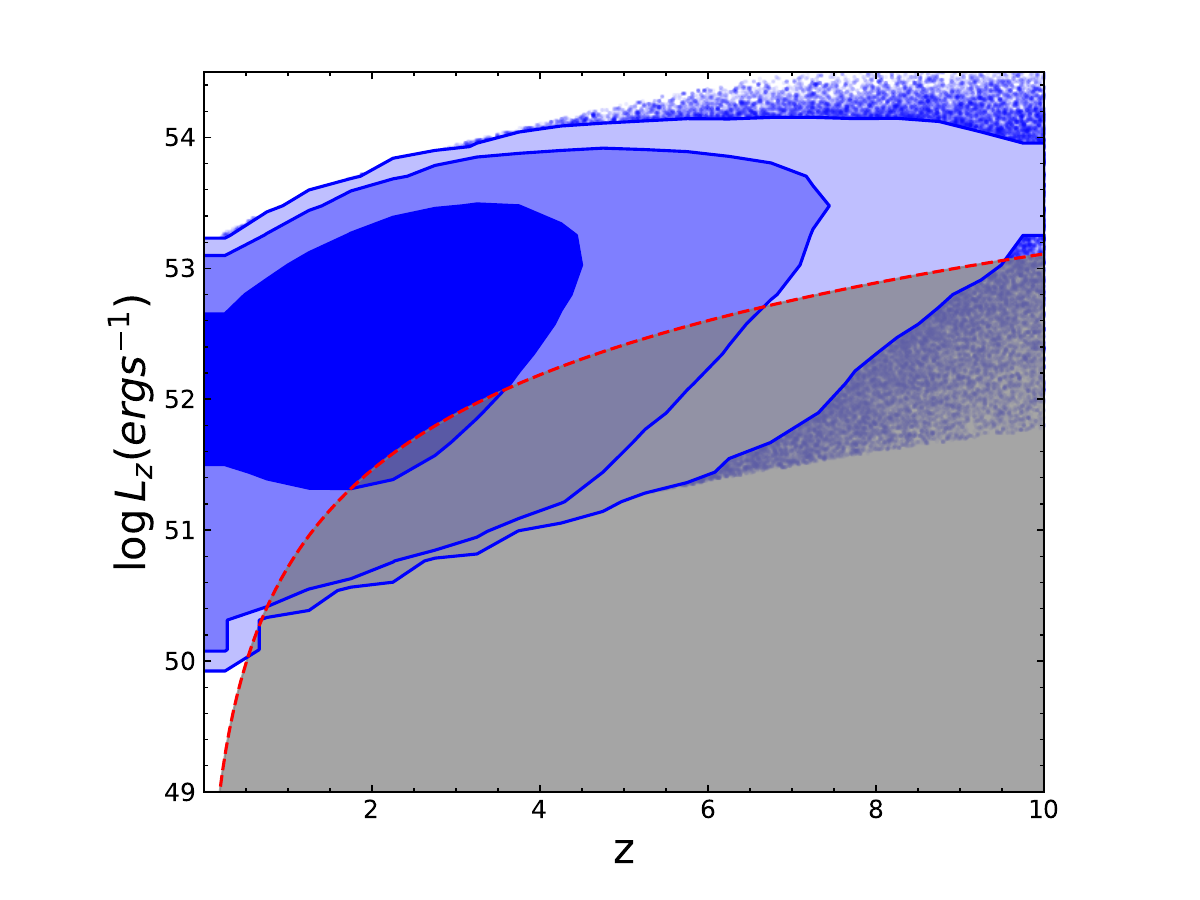}
	\includegraphics[scale=0.45]{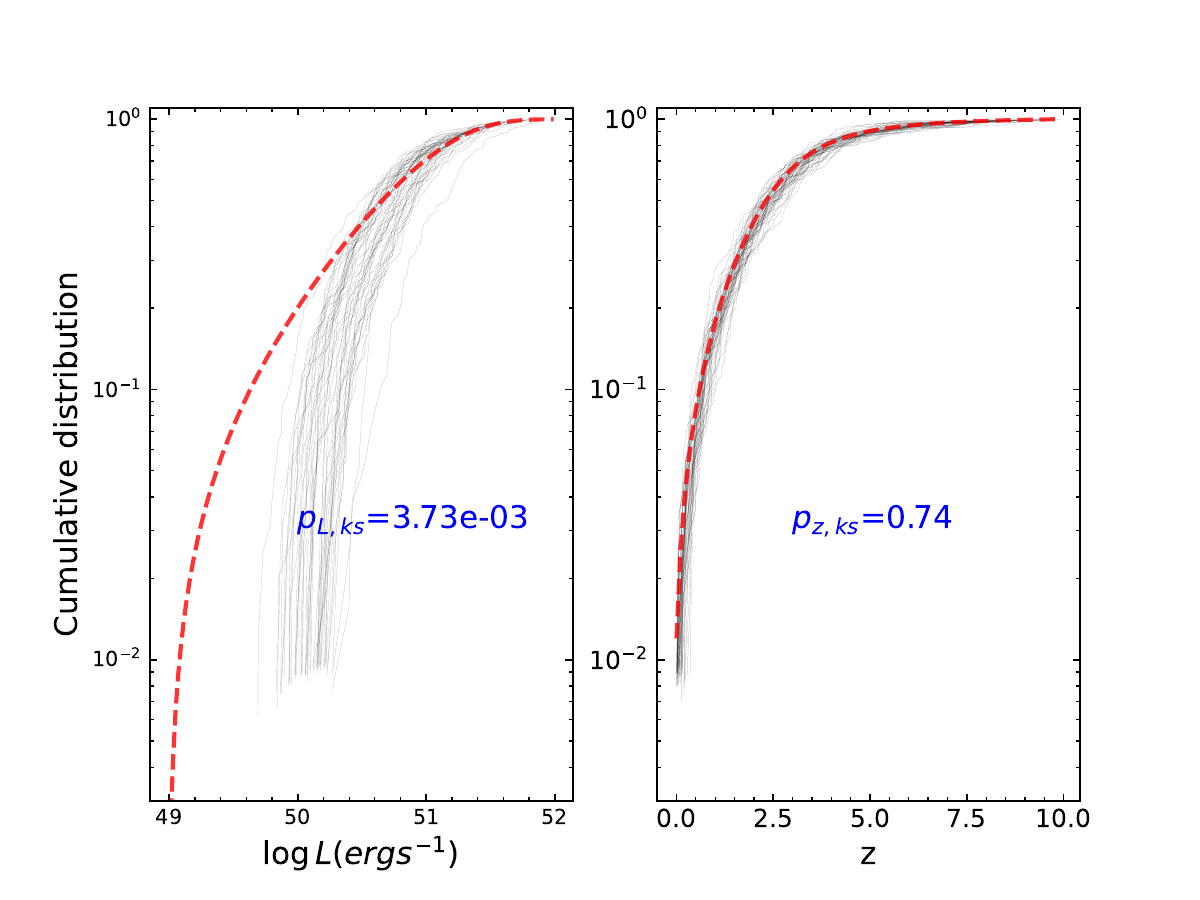}
	\includegraphics[scale=0.45]{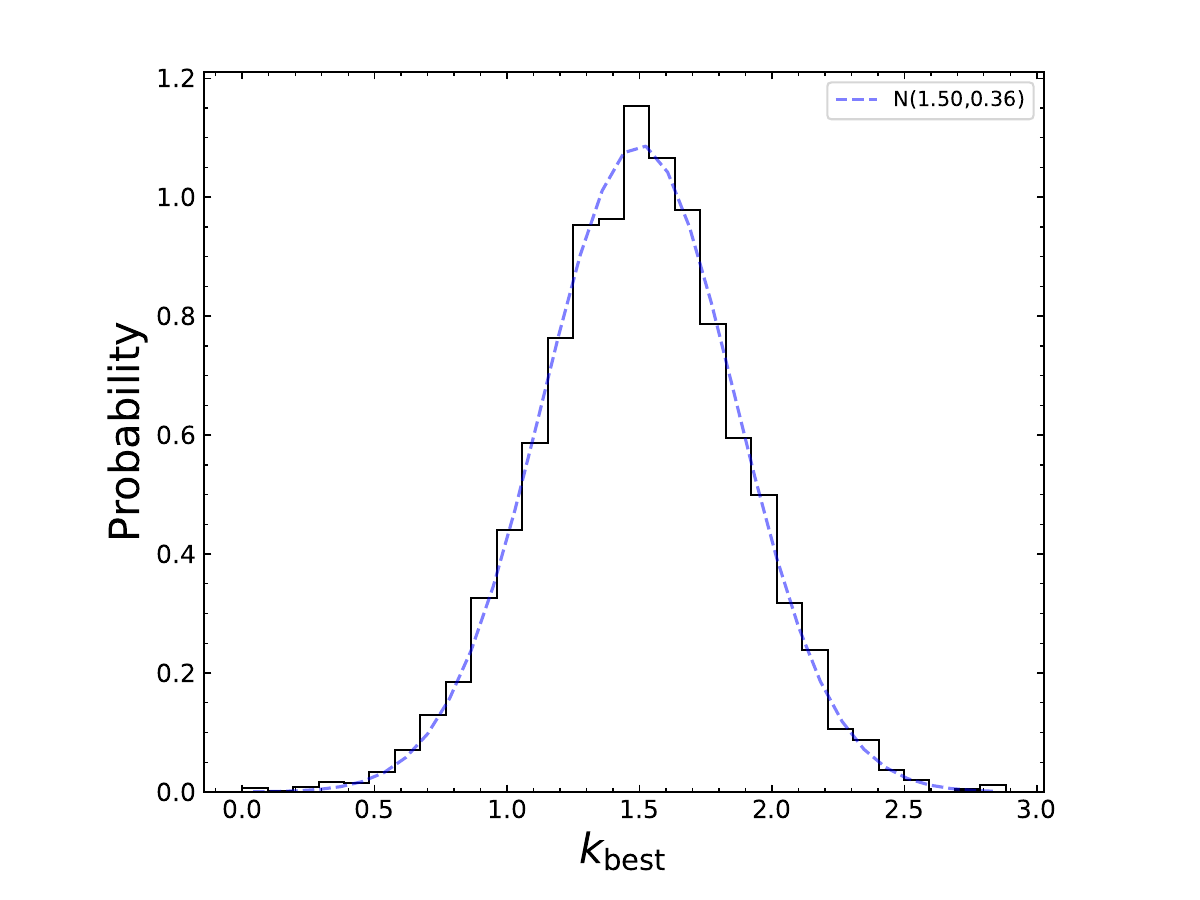}
	\caption{Demonstrations of biased reconstruction of the underlying parent population from truncated data under misconfigured transformation function, $g(z)$. Upper panel: the same as the upper of Fig. (\ref{Fig:LFz}), but for the intrinsic luminosity evolving with redshift by $L_{z}=L(3+z)^{2.5}$. Center panel: the reconstructed luminosity and redshift distributions from the truncated data (bright area in the upper panel), are compared to their corresponding population distributions (the red dashed lines), respectively, in which the luminosity evolution, $g(z)=(1+z)^{k}$ is assumed. The distribution of the $k_{\rm best}$ at $\tau=0$ is shown in the lower panel. The symbols are the same as the lower panels of Figs. of (\ref{Fig:LF}) and (\ref{Fig:LFz}).}
	\label{Fig:LFz2}
\end{figure}

However, this is not the case when the detailed information of intrinsic luminosity evolutions is not known. As shown in Fig  (\ref{Fig:LFz2}) is an instance of the case, in which, we assume that the intrinsic luminosity evolves with redshift by the law of $L_{z}=L g_{\rm int}(z)$, where $g_{\rm int}=(3+z)^{2.5}$, is taken to parameterize its intrinsic luminosity evolution. Then, a bias reverse transformation function, such as, $g(z)=(1+z)^{k}$, is adopted (The same as that usually adopted by some authors in the astronomical context) to reconstruct their underlying parent population  as done in Fig (\ref{Fig:LFz}).  We find that the $k_{\rm best}$ at $\tau=0$ also obeys a Gaussian distribution. Then a Gaussian function is employed to fit. This yields a value of $\bar{k}_{\rm best}=1.50\pm0.36$, which differs significantly from its intrinsic evolution index. In this case, although its redshift distribution can be unbiasedly recovered, the underlying luminosity distribution can not. Certainly, in this instance, if the reverse transformation function, $g(z)=(3+z)^{k}$, is adopted, their underlying parent population could also unbiasedly recovered.

The fact turns out that, in the reconstruction of an intrinsic luminosity function, if using a misconfigured transformation function, one does not unbiasedly recover its underlying parent population, though such a transformation does produce an uncorrelated truncated sample.
%%%%%%%%%%%%%%%%%%%%%%%%%%%%%%%%%%%%%%%%%%%%%%%%%%

\section{SUMMARY and DISCUSSION } \label{sec:diss}

In this work, we use Monte Carlo simulation to explore what factors would impact the $\tau$ statistic, and that how one could unbiasedly recover the underlying parent population from a truncated sample based on Lynden-Bell's $c^{-}$ method. Our main results are as follows.

1. According to \cite{1992ApJ...399..345E}, the $\tau$ statistic, measured by Equation \ref{eq:tau}, always follows a standard normal distribution (see Fig.\ref{taus}) under the premise that the underlying bivariate variables are uncorrelated regardless of adopted selection functions. Under the condition, an uncorrelated bivariate population distribution could always be unbiasedly recovered from a truncated sample with Lynden-Bell's nonparametric $c^{-}$ method (Please refer to \cite{2020sdmm.book.....I}).

2. On the contrary, when an observable sample comes from an underlying correlated bivariate population distribution, the $\tau$ statistic no longer obeys a standard normal distribution, but a normal distribution with both its average and standard variance changing with the Pearson’s coefficient ($r$) of the population, the observable sample size ($N_{\rm Tr}$), as well as different selection functions (see Fig.\ref{tau-N}), which indicates that, in this situation, the origin of the $\tau$ statistic is a complex combination of multiple factors. In this case, it is very difficult to unbiasedly recover the underlying population from a truncated sample with Lynden-Bell's $c^{-}$ method by the transformation, unless the detailed information of the intrinsic correlation is known accurately in advance, then its corresponding reverse transformation is applied in the constructions (see Fig. \ref{Fig:LF}).

3. In practice, the luminosity evolution form, $g(z)=(1+z)^{k}$, derived from a truncated sample with the $\tau$ statistical method, does not necessarily represent its underlying luminosity evolution.  

By applying the transformation function, $g(z)=(1+z)^{k}$, to several GRBs samples, \cite{2021MNRAS.504.4192B} found that the resulting $k$ is sensitive to be adopted detection thresholds. This fact may indicate that these GRBs samples may likely come from an intrinsic correlated population, according to what we find in Fig. (\ref{tau-N}). If so, it is extremely difficult to get the detailed information of the intrinsic correlation between the luminosity and redshift of GRBs. 

Again, with the transformation method, some authors \citep{2015ApJS..218...13Y,2015ApJ...806...44P,2019MNRAS.488.5823L} found a low-redshift excess in the formation rate of GRBs , whereas others did not \citep{2016A&A...587A..40P,2017ApJ...850..161T}. Whether the low-redshift excess results from an intrinsic physics?  or from an inappropriate transformation method? or is biased by the sample size and completeness.  It needs to be further investigated with a larger complete observed sample in the future.

\section*{Acknowledgements}
This work is supported by the National Natural Science Foundation of China (grant No. 12133003). 
	
\vspace{5mm}

\section*{Data Availability}
The data underlying this article will be shared on reasonable request
to the corresponding author.

	
\bibliographystyle{mnras}
\bibliography{example} % if your bibtex file is called example.bib  
%\bibliography{References}

%\begin{thebibliography}{99}
%\end{thebibliography}

%%%%%%%%%%%%%%%%%%%%%%%%%%%%%%%%%%%%%%%%%%%%%%%%%%

%%%%%%%%%%%%%%%%% APPENDICES %%%%%%%%%%%%%%%%%%%%%

%%%%%%%%%%%%%%%%%%%%%%%%%%%%%%%%%%%%%%%%%%%%%%%%%%

% Don't change these lines
\bsp	% typesetting comment
\label{lastpage}
\end{document}